\documentclass[journal=nalefd,manuscript=article]{achemso}
\usepackage{chemformula} 
\usepackage[T1]{fontenc} 
\usepackage{bm}
\usepackage[normalem]{ulem}
\usepackage{xcolor}
\usepackage{fancyhdr,graphicx}
\usepackage{bm}
\usepackage{color}
\usepackage{soul}
\usepackage{amsfonts}
\usepackage{times}
\usepackage{amsmath}
\usepackage[colorlinks=true, linkcolor=blue, citecolor=blue, urlcolor=blue]{hyperref}
\setcitestyle{journalcolor= black}
\usepackage{wrapfig}
\usepackage{amsmath} 

\usepackage{easyReview}
\usepackage{url}

\usepackage{babel}
\usepackage[normalem]{ulem}
\usepackage{units}
\usepackage{amssymb}
\usepackage{soul}
\usepackage{fancyhdr,graphicx}
\usepackage{bm}
\usepackage{parskip} 
\usepackage{natbib}
\usepackage{wrapfig}
\usepackage{easyReview}
\usepackage{color}
\usepackage{nameref}
\usepackage{todonotes}
\def\be{\begin{equation}}
\def\ee{\end{equation}}
\def \bea{\begin{eqnarray}}
\def \eea{\end{eqnarray}}
\def \nn{\nonumber}

\title{Giant gate-controlled room temperature odd-parity magnetoresistance in magnetized bilayer graphene}
\author{Divya Sahani$^1$, Sunit Das$^{2}$, Kenji Watanabe$^3$, Takashi Taniguchi$^4$, Amit Agarwal$^{2*}$, and Aveek Bid$^1$}
\email{{amitag@iitk.ac.in, aveek@iisc.ac.in}}
\affiliation{$^1$Department of Physics, Indian Institute of Science, Bangalore 560012, India \\
	$^2$Department of Physics, Indian Institute of Technology Kanpur, Kanpur-208016, India\\
	$^3$Research Center for Functional Materials, National Institute for Materials Science, 1-1 Namiki, Tsukuba 305-0044, Japan \\
	$^4$International Center for Materials Nanoarchitectonics, National Institute for Materials Science, 1-1 Namiki, Tsukuba 305-0044, Japan \\
}


\begin{document}

\begin{abstract}
Magnetotransport measurements are crucial for understanding the Fermi surface properties, magnetism, and topology in quantum materials. Here, we report the discovery of giant room temperature odd-parity magnetoresistance (OMR) in a bilayer graphene (BLG) heterostructure interfaced with \ch{Cr_2Ge_2Te_6} (CGT). Using magnetotransport measurements, we demonstrate that the BLG/CGT heterostructure exhibits a significant antisymmetric longitudinal magnetoresistance, indicative of intrinsic time-reversal symmetry (TRS) breaking in the system. We show that the OMR is tunable via electrostatic gating. Additionally, the OMR is pronounced near the band edges and diminishes with increasing charge carrier density in graphene. Our theoretical analysis reveals that this phenomenon arises from the coupling of the out-of-plane components of Berry curvature and orbital magnetic moment to the applied magnetic field in a TRS-broken system. Our findings establish OMR as a significant probe for TRS breaking in quantum materials in which the crystal symmetries preclude the appearance of anomalous Hall effect.
\end{abstract}

\maketitle


\section{Introduction}
\sloppy
Magnetotransport measurements provide a powerful tool for exploring various material properties, such as symmetry, topology, and magnetism~\cite{Lu2017}. The discovery of giant magnetoresistance~\cite{Chazekas_prl88} and tunneling magnetoresistance~\cite{Yuasa_nm04} has enabled the development of highly efficient, non-volatile storage and memory devices~\cite{Katti}. In high magnetic fields, metals typically exhibit oscillating magnetoresistance, known as the Shubnikov-de Haas effect, which is routinely used to measure the Fermi surface properties~\cite{glazman_fermiology23, das2024nonlinearlandaufandiagram}.
In the low magnetic field regime, the longitudinal magnetoresistance usually becomes a monotonic, even function of the externally applied magnetic field ($B$). The magnetoresistance follows the Onsager's reciprocity relation \({R}_{ab}(B) = {R}_{ba}(-B)\)~\cite{onsager1931reciprocal, Mazur_53}, with $\{a,b\}=\{x,y\}$. This relation implies that the off-diagonal resistance (\(a \neq b\)) can have both symmetric $[{ R}_{ab}(B) = {R}_{ab}(-B)]$ and antisymmetric $[{ R}_{ab}(B) = -{R}_{ab}(-B)]$ components, while the longitudinal magnetoresistance can only be symmetric $[{ R}_{aa}(B) = {R}_{aa}(-B)]$ or an even function of $B$.

This traditional understanding of longitudinal magnetoresistance fails when the system intrinsically breaks time-reversal symmetry (TRS) without an external magnetic field. A generalized Onsager's reciprocal relation still holds, provided all the TRS breaking fields are reversed~\cite{Mazur_53, Buttiker_prb12}: For example, ${ R}_{ab}(B, M) = {R}_{ba}(-B, -M)$, with $M$ being the intrinsic TRS breaking field, typically associated with magnetization or magnetic exchange term. In such cases, the symmetry restriction on the longitudinal magnetoresistance gets modified, and an antisymmetric, linear-$B$ longitudinal magnetoresistance with odd parity can be finite~\cite{Cortijo_prb16, Tiwari_prb17, Das_prb19_electrical, tobius_prb20, Nui_prb20, Zyuzin_prb21, Wu_prl21, lahiri_prb22, Orenstein_23linear, ghorai_24planar, albarakati2019antisymmetric, Wang_NC20, niu2021antisymmetric, jiang2021chirality}. Thus, experimental observation of antisymmetric in $B$, or odd-parity linear magnetoresistance (OMR), can evidence intrinsic TRS breaking.

The intrinsic TRS breaking of a system can also be probed by the Berry curvature-driven anomalous Hall effect (AHE) in the system. However, crystalline symmetry restrictions sometimes force the AHE to vanish in specific TRS broken systems while retaining a finite odd parity longitudinal magnetoresistance. Examples of this include TRS breaking systems that preserve in-plane mirror symmetries like ${\cal M}_x$ or ${\cal M}_y$~\cite{Orenstein_23linear}. Additionally, if the crystalline symmetry-breaking parameter is small, the AHE becomes too small to be detected. Antisymmetric or odd-parity longitudinal magnetoresistance (OMR) provides a reliable probe for TRS breaking in such materials.

In this Letter, we report a giant OMR in a heterostructure of bilayer graphene (BLG) coupled to bilayer \ch{Cr_2Ge_2Te_6} (CGT), persisting till room temperature and beyond. Our measurements establish that the OMR is large near the band edges and decreases with increasing charge carrier number density in graphene. Remarkably, other signatures of the effect of exchange interaction on electrical transport properties, namely hysteresis in magnetoresistance and the anomalous Hall effect, are conspicuously absent. Interestingly, we find that the sign of the OMR switches when the voltage terminal is changed to the opposite edges~\cite{takiguchi2022giant}.
Our theoretical analysis reveals this unique OMR phenomenon stems from the out-of-plane components of the Berry curvature and orbital magnetic moment in a time-reversal symmetry broken system \cite{PhysRevB.103.125432,PhysRevLett.129.227401,Adak_2024}. The proximity to magnetic \ch{Cr_2Ge_2Te_6} provides the magnetic exchange coupling in BLG, which is essential for this phenomenon. Our discovery of a giant gate-tunable room temperature OMR motivates further exploration of novel magnetotransport phenomena and potential technological applications such as magnetic sensors and memory devices in magnetized van-der-Waals materials.

\section{Results}

Heterostructures of CGT/BLG/hBN were fabricated on $\mathrm{Si}^{++}$\ch{/SiO2} substrates using the dry transfer technique~\cite{pizzocchero2016hot} (Fig.~\ref{fig:fig1}(a)). The thickness of the CGT was between two and three layers.
One-dimensional electrical contacts were created using electron beam lithography followed by dry etching
and metallization using Cr/Pd/Au (3~nm/13~nm/55~nm); see Supplementary Information Sec. S1 for fabrication details. A thick layer of graphite ($\approx 20$ nm thick) was used as the top-gate electrode, while the heavily doped Si acted as the bottom-gate electrode. The device was dry etched into a Hall bar. Inset of Fig.~\ref{fig:fig1}(b) shows a schematic of the device with the numbered electrical contacts. An optical image of the device is shown in the inset of Fig.~\ref{fig:fig1}(c). In this Letter, we present the data from a device labeled \textbf{D1}; data for other devices are provided in Supplementary Information. We compare the data with those from a heterostructure of \ch{hBN/BLG/hBN} (labeled \textbf{D2}).

Electrical transport measurements were done using a low-frequency lock-in detection technique at a bias current \ch{10 nA} in a variable temperature cryostat. CGT is an electrical insulator at low temperatures~\cite{verzhbitskiy2020controlling}. Consequently, electrical transport only takes place through the graphene channel. The dual-gated geometry allows independent control of the charge carrier density $n$ and the vertical displacement field $D$ {through $n=[(C_{tg}V_{tg}+C_{bg}V_{bg})/e+n_{0}]$ and $D=[(C_{bg}V_{bg}-C_{tg}V_{tg})/{2}]$ across the device. Here, $C_{tg}$ and $C_{bg}$ are the top-gate and back-gate capacitance, respectively. Their values are determined from quantum Hall measurements. $V_{bg}$ and $V_{tg}$ are the back-gate and top-gate voltages.  $n_{0}$ is the residual charge carrier number density due to doping from CGT  and channel impurities. The variation of the longitudinal resistance with gate voltage shows that the charge neutrality point of the BLG is shifted to a positive gate voltage (Sec.~S2 of Supplementary Information), indicating that the BLG is hole-doped due to charge transfer at the interface from CGT~\cite{tseng2022gate,tenasini2022band,wang2022quantum,chau2022two,yang2023unconventional}. Hall resistance confirm that the hole-doping is ${n_0 = -9.6\times 10^{15}}$~$ \mathrm{m^{-2}}$ (Fig. S1 of Supplementary Information). Such large charge transfer between CGT and graphene has been noted in previous studies~\cite{tseng2022gate,yang2023unconventional}. Here, we establish that the charge transfer magnitude increases with increasing CGT thickness; see Fig.~S1 of Supplementary Information. Contrast this with the $R$-$V_g$ plot of device \textbf{D2}---the charge neutrality point is at $V_g = 0.04$ indicating ${n_0 = 1.53\times 10^{14}}$~$ \mathrm{m^{-2}}$ (Sec. S2 of Supplementary Information).
\subsection{Primary observations}

In Fig.~\ref{fig:fig1}(c), the solid blue line represents the plot of the longitudinal magnetoresistance $ R_{14,65}(B)$ as a function of the magnetic field $B$ measured at ${n = -6.28 \times 10^{16}}$~$ \mathrm{m^{-2}}$. The notation ${ R}_{ij,kl}(B)$ indicates that the current is driven between the probes $i$ and $j$ in the presence of a perpendicular magnetic field $B$. At the same time, the potential difference is measured between the probes $k$ and $l$. Surprisingly, the magnetoresistance has contributions which are asymmetric in the $B$-field. In contrast, the longitudinal magnetoresistance measured for the pristine bilayer graphene encapsulated between hBN layers, \textbf{D2}, is completely symmetric in the $B$-field (Fig.~\ref{fig:fig1}(c) right-axis; solid red line).

Figure~\ref{fig:fig1}(d) is a plot of the symmetric (even-parity) component of longitudinal magnetoresistance ${{ R}^{\rm S}_{14,65}={[ {R}_{14,65}(B)+ {R}_{14,65}(-B)}]/2}$. The OMR
${{ R}^{\rm AS}_{14,65}= [{R}_{14,65}(B)- {R}_{14,65}(-B)]/2}$ is presented in Fig.~\ref{fig:fig1}(e).
We note two essential characteristics of the OMR. Firstly, its change with the magnetic field is enormous, approaching $40 \%$ at 14~T~(Fig.~S3 of Supplementary Information). This value is the highest reported to date~\cite{Nui_prb20,takiguchi2022giant}, see Supplementary Information Table~S1. Secondly, the OMR is linearly dependent on the ${B}$ field (onto which quantum oscillations are superimposed). Note that we do not observe any hysteresis in the magnetoresistance of the BLG/CGT devices down to 2~K (see Sec.~S3 of the Supplementary Information for data).

We observe a strong electrostatic gate control on the magnitude of OMR. In Fig.~\ref{fig:fig2}(a),  ${R^{\rm AS}_{14,65}}$  is plotted for three different charge carrier densities ${n= -6 \times 10^{16}~ \mathrm{m^{-2}}}$ (green curve), ${n= -2 \times 10^{16}~ \mathrm{m^{-2}}}$ (black curve) and ${n= -1 \times 10^{16}~ \mathrm{m^{-2}}}$ (orange curve). The OMR decreases steadily with increasing career density. We quantify the magnitude of the OMR using the parameter $\alpha ={d}{ R}^{\rm AS}_{ij,kl}/{d}B$ and find that the magnitude of $\alpha$ decreases rapidly with increasing $|n|$ (Fig.~\ref{fig:fig2}(c)). This observation establishes that the OMR is maximum at the band edges, providing the first clue about its origin.

Fig.~\ref{fig:fig2}(b) shows the plot of ${{R}^{\rm AS}_{14,65}}$ versus ${B}$ at different temperatures ranging between $2-360$~K measured at a hole number density $|n| = 2\times 10^{16}$~$\mathrm{m^{-2}}$. As expected, the SdH oscillations disappear at higher temperatures. Surprisingly, the amplitude of ${{R}^{\rm AS}_{14,65}}$ remains almost unchanged till about $T = 250$~K before beginning to fall precipitously, finally disappearing at $T \sim 360$~K (see Fig.~\ref{fig:fig2}(d)). The observation of a gate-tunable giant OMR at room temperature and beyond is the central result of this Letter.

Before we comprehensively explain the origin of the OMR, we rule out two artifacts that can lead to an OMR. A trivial origin of OMR in semiconductors is inhomogeneities in the device where a part of the transverse Hall voltage is reflected along the longitudinal direction of current flow~\cite{khouri2016linear,stormer1992strikingly, PhysRevResearch.5.L032046}. In Supplementary Information Sec.~S5, we rule out this interpretation. Another possible origin can be intermixing from the Hall signal. We rule this out by noting that $\alpha$ has the same sign for both electron and hole-doping of graphene while $R_{xy}$ changes its sign between these two regimes (see Sec.~S6 of Supplementary Information).

\subsection{Theoretical model}

The OMR in TRS broken systems can be qualitatively understood from a semi-classical transport framework, including the co-existence of topological bands and magnetic order. Using the Boltzmann transport formalism, we show that the linear-in-$B$ longitudinal conductivity arises from the orbital magnetic moment and band geometric quantities such as the Berry curvature. Assuming the electric field due to the source-drain bias is in the $x$-direction, we obtain the linear longitudinal magnetoconductivity to be (see Sec.~S7 of Supplementary Information for calculation details)
\bea \label{sigma_B}
\sigma_{xx}^{\rm AS}(B) =&& - 2 e^2 \tau  \int_{p,{\bm k}} {v}_{x}^m {v}_{x}^0 ~\partial_{\varepsilon} f_p^0 + \frac{e^3\tau}{\hbar} \int_{p,{\bm k}} ({\bm \Omega}\cdot {\bm B}) {v}_{x}^0 { v}_{x}^0 ~\partial_{\varepsilon} f_p^0 \nn \\
&&+ e^2 \tau \int_{p,{\bm k}} { v}_{x}^0 {v}_{x}^0  \left[ \partial_{\varepsilon}({\bm m}\cdot {\bm B}) ~\partial_{\varepsilon} f_p^0 + ({\bm m}\cdot {\bm B}) ~\partial^2_{\varepsilon} f_p^0 \right].
\eea
For brevity, we denote $\int_{p,{\bm k}} \equiv \sum_p \int \frac{d^2{k}}{(2\pi)^2}$ and we do not explicitly mention the band index $p$ in the physical quantities. In Eq.~\eqref{sigma_B}, $ {\bm v}^0 ={\bm \nabla}_{\bm k}\varepsilon_0/\hbar$ is band velocity, and $ {\bm v}^m ={\bm \nabla}_{\bm k}(-{\bm m}\cdot {\bm B})/\hbar$ is the correction to band velocity arising from the coupling of the orbital magnetic moment $\bm m$ to the magnetic field. $\tau$ is the scattering time, and $f_p^0$ is the equilibrium Fermi-Dirac distribution function. Interestingly, $\bf{\Omega}$, ${\bm m}$ and $\bm{v}^m$ are odd functions of ${\bm k}$ in presence of TRS. Consequently, all the integrals in Eq.~\eqref{sigma_B} become odd function of $k$ and vanish in systems preserving TRS. Hence, the Berry curvature and orbital magnetic moment-induced linear-$B$ longitudinal magnetoconductivity can only be finite in a TRS-broken system. Several other theoretical models such as the Parish-Littlewood model~\cite{Parish2003}, Abrikosov quantum theory~\cite{Abrikosov_prb98, Abrikosov2000}, guiding center diffusion mechanism~\cite{Song_prb15}, and other mechanisms~\cite{Wang_prb20,takiguchi2022giant, Sinchenko_prb17,Wu_prb20, Feng_pnas19} have been proposed to account for linear magnetoresistance. However, we have carefully examined each of these theories in Sec.~S8 of our Supplementary Information, and find that none of these are relevant to our system.

In Fig.~\ref{fig:fig2}(e), we present the calculated variation of the antisymmetric magneto-conductivity ($\sigma^{\rm AS}_{xx} \propto \rho^{\rm AS}_{xx} \propto B$) with the hole density $|n|$ for the model Hamiltonian ({see Sec.~S7 of Supplementary Information) of bilayer graphene proximitized with a single layer of CGT~\cite{Fabian_prb21}. We observe that $\sigma^{\rm AS}_{xx}$ (and, consequently, the antisymmetric part of the resistivity, $\rho^{\rm AS}_{xx} \approx \sigma^{\rm AS}_{xx} \rho^{\rm S}_{xx}/\sigma^{\rm S}_{xx}$) decreases with increasing number density. Here, $\rho_{xx}^{\rm S}$ and $\sigma_{xx}^{\rm S}$ are the symmetric part of the longitudinal resistivity and the conductivity. See Supplementary Information Sec.~S7 for more details. This qualitatively captures our experimental finding of $\alpha$ decreasing on increasing $|n|$, shown in Fig.~\ref{fig:fig2}(c).

\subsection{Characteristics of the OMR}

While both bulk and few-layer CGT have been shown to exhibit perpendicular magnetic anisotropy~\cite{wu2022van}, the magnetic properties of two- or three-layer thick CGT are unexplored. The $T$-dependence of OMR hints at a large enhancement of the magnetic ordering temperature in bilayer CGT on bilayer graphene heterostructure. This is similar to the dramatic Curie temperature enhancement observed in ultra-thin 2D materials due to substrate effect~\cite{PhysRevLett.99.117205, https://doi.org/10.1002/adma.202002032, doi:10.1021/acsami.8b04289,Zhang_molecules23}. To explain the observed $T$-dependence of $\alpha$, we make an implicit assumption that the OMR is proportional to the magnetization in the system~\footnote{\label{fn:note1}Our assumption stems from the physical intuition that the response parameter ($\alpha$) resulting from TRS breaking should be proportional to the TRS breaking parameter ($M$). We have also checked this explicitly for the analytically tractable example of gapped graphene. However, concrete and general proof of this statement is still lacking}. As $T$ approaches the critical temperature $T_c$, the magnetization of the system should vanish as $M=M_0 (1-T_c/T)^\beta$ ($\beta$ being the critical exponent quantifying the dependence of $M$ on $T$). The intrinsic time-reversal symmetry of the system is restored for $T>T_c$ in the paramagnetic phase where $M = 0$. Consequently, the antisymmetric part of the magnetoconductivity should become zero beyond the critical temperature. In the spirit of the above argument, we fit the data in Fig.~\ref{fig:fig2}(d) with the relation $\alpha=\alpha_0 (1-T_c/T)^\beta$; the fits yield $T_c\sim 360$~K and $\beta \sim 0.198$. This value of $\beta$ matches very well with that previously extracted from magnetization measurements CGT; $\beta = 0.240 \pm 0.006$~\cite{PhysRevB.95.245212}. We show a plot of the calculated antisymmetric magnetoconductivity $\sigma_{xx}^{\rm AS}$ versus $T$ in Fig.~\ref{fig:fig2}(f) using these experimentally obtained values of $T_c$ and $\beta$ for the proximity induced magnetization. The resulting dependence of calculated $\sigma_{xx}^{\rm AS}$ on $T$ captures the experimental data exceptionally well.

Fig.~\ref{fig:fig3}(a) is the plot of ${R^{\rm AS}_{14,65}}$ for different values of bias current $I$ measured at ${2}$~K; the data have been vertically offset for clarity. The magnitude of OMR is independent of the strength of the driving current, as is apparent from the plot of $\alpha$ versus $I$ (inset of  Fig.~\ref{fig:fig3}(a)). Furthermore, Fig.~\ref{fig:fig3}(b) compares ${R^{\rm AS}_{14,65}}$ (the current is driven from contact 1 with contact 4 grounded) and ${R^{\rm AS}_{41,56}}$ (the current is driven from contact 4  with contact 1 grounded); these two plots overlap perfectly. These observations establish the OMR as a reciprocal, linear response transport phenomenon with an origin distinct from recent reports where the OMR had a strong current dependence~\cite{takiguchi2022giant}.

In 2D materials, only the out-of-plane component of the orbital magnetic moment ($m_z$) and Berry curvature ($\Omega_z$) are finite. The $m_z$ couples to the out-of-plane magnetic field and modifies the band energy via Zeeman-like coupling: $\varepsilon\to (\varepsilon_0 -{\bm m}\cdot {\bm B})$. Similarly, a term $\propto {\bm \Omega}\cdot {\bm B}$ modifies the phase space volume element (see Eq.~\eqref{sigma_B} and Sec.~S7 of Supplementary Information). Thus, the in-plane component of the magnetic field can play no role in the observed antisymmetric longitudinal conductivity (or resistivity) induced by the $m_z$ and $\Omega_z$. We verified this assertion by measuring OMR in a tilted magnetic field. For these measurements, the device was rotated with respect to the magnet's axis. One can see from  Fig.~\ref{fig:fig3}(c) that for a given value of the perpendicular component of the magnetic field, $B_\perp$, the OMR is independent of the angle between the magnetic field and the perpendicular to the plane of the device, $\theta$ (and consequently of the parallel component of the magnetic field, $B_\parallel$). This independence of OMR on $B_\parallel$ further validates the consistency of our theoretical understanding with the experimental observation. It also rules out other possible spin-based mechanisms~\cite{Wang_prb20, takiguchi2022giant} of generating OMR via the spin-Zeeman coupling (see Sec.~S8 of Supplementary Information for more details).

In Fig.~\ref{fig:fig4}(a), we plot the OMR measured for the bottom (black curve, ${R_{14,65}}$) and the top edges (red curve, ${R^{\rm AS}_{14,23}}$) of the device. The OMR switches sign between the two edges. Given that the direction of the current and the magnetic field in these two measurements remain identical, the flipping of the sign of OMR between the two edges indicates the presence of trivial 1-D edge modes in addition to the 2-D bulk mode~\cite{takiguchi2022giant}. Such trivial non-topological edges have been observed in etched graphene devices and attributed to charge accumulation at the edges~\cite{aharon2021long}. We do not completely understand the origin of these trivial edge states in our CGT/BLG devices. On the other hand, as detailed in Ref.~\cite{takiguchi2022giant}, the 2-probe longitudinal resistance ${R_{14,65}}$ must be $B$-field symmetric; Fig.~\ref{fig:fig4}(b) establishes that this indeed is the case.

\section{Conclusions}

In proximity to a magnetic insulator, bilayer graphene can exhibit the anomalous Hall effect and exchange splitting during field cooling experiments. However, the exchange interaction weakens for ultra-thin ferromagnet layers, making anomalous Hall signals challenging to detect. Additionally, specific crystalline symmetries can also force the anomalous Hall signals to vanish. Our use of odd-parity magnetoresistance (OMR) as a probe provides a reliable alternative method to detect broken time-reversal symmetry in these materials.

Observing OMR in bilayer graphene (BLG)/Cr$_2$Ge$_2$Te$_6$ (CGT) heterostructures at room temperature is particularly exciting. The most viable mechanism for this phenomenon in the BLG/CGT system is the interplay between the band geometric properties of graphene and the magnetic exchange interaction induced by CGT. Coupled with our observation of room temperature OMR, this implies that the magnetic ordering in bilayer CGT on graphene persists up to room temperature. Given that the ferromagnetic Curie temperature of bulk CGT is around 65 K, our finding that bilayer CGT/BLG heterostructure possesses a finite magnetic moment at room temperature is very significant. Similar behavior has also been reported in other graphene-based heterostructures~\cite{Zhang_molecules23, Chang_24_atomic}. Our observation of OMR beyond room temperature necessitates further theoretical and experimental investigations into the magnetic and topological properties of the BLG/CGT heterostructure. Specifically, the existence of room temperature magnetization has to be confirmed through independent magnetization measurements. The possibility of room temperature magnetization in van-der-Waals heterostructure is very remarkable and holds immense  potential for practical technological applications.




 \begin{acknowledgement}
A.B. acknowledges funding from the Department of Science \& Technology FIST program and the U.S. Army DEVCOM Indo-Pacific (Project number: FA5209   22P0166). A.A. acknowledges the Science and Engineering Research Board for Project No. MTR/2019/001520, and the Department of Science and Technology for Project No. DST/NM/TUE/QM-6/2019(G)-IIT Kanpur of the Government of India for funding. K.W. and T.T. acknowledge support from JSPS KAKENHI (Grant Numbers 19H05790, 20H00354, and 21H05233). S.D. acknowledges the Ministry of Education, Government of India, for funding support through the Prime Minister's Research Fellowship program.\end{acknowledgement}


\section{Author Contributions}

D.S. and A.B. conceptualized the study, performed the measurements, and analyzed the data.   A.A. and S.D. performed the theoretical analysis. K.W. and T.T. grew the hBN single crystals. All the authors contributed to preparing the manuscript.

\section{Data availability}

\noindent The authors declare that the data supporting the findings of this study are available within the main text and its Supplementary Information. Other relevant data are available from the corresponding author upon reasonable request.\\

\section{Competing interests}

\noindent The authors declare no Competing Financial or Non-Financial Interests.\\

	\begin{suppinfo}

Supporting information contains detailed discussions of the (S1) device fabrication, (S2) gate voltage dependence of the longitudinal resistance, (S3) comparison of zero field-cooled and field-cooled magnetoresistance, (S4) large \% change in antisymmetric magnetoresistance, (S5) the possible trivial origin of OMR: carrier density inhomogeneity picture, (S6) ruling out Hall intermixing as the origin of OMR: analysis of Hall and OMR for holes and electrons, (S7) band geometry induced antisymmetric magnetoresistance, and (S8) discussion on different theoretical models for OMR.

	\end{suppinfo}
	\clearpage

\begin{figure*}[t!]
	\begin{center}
			\includegraphics[width=\columnwidth]{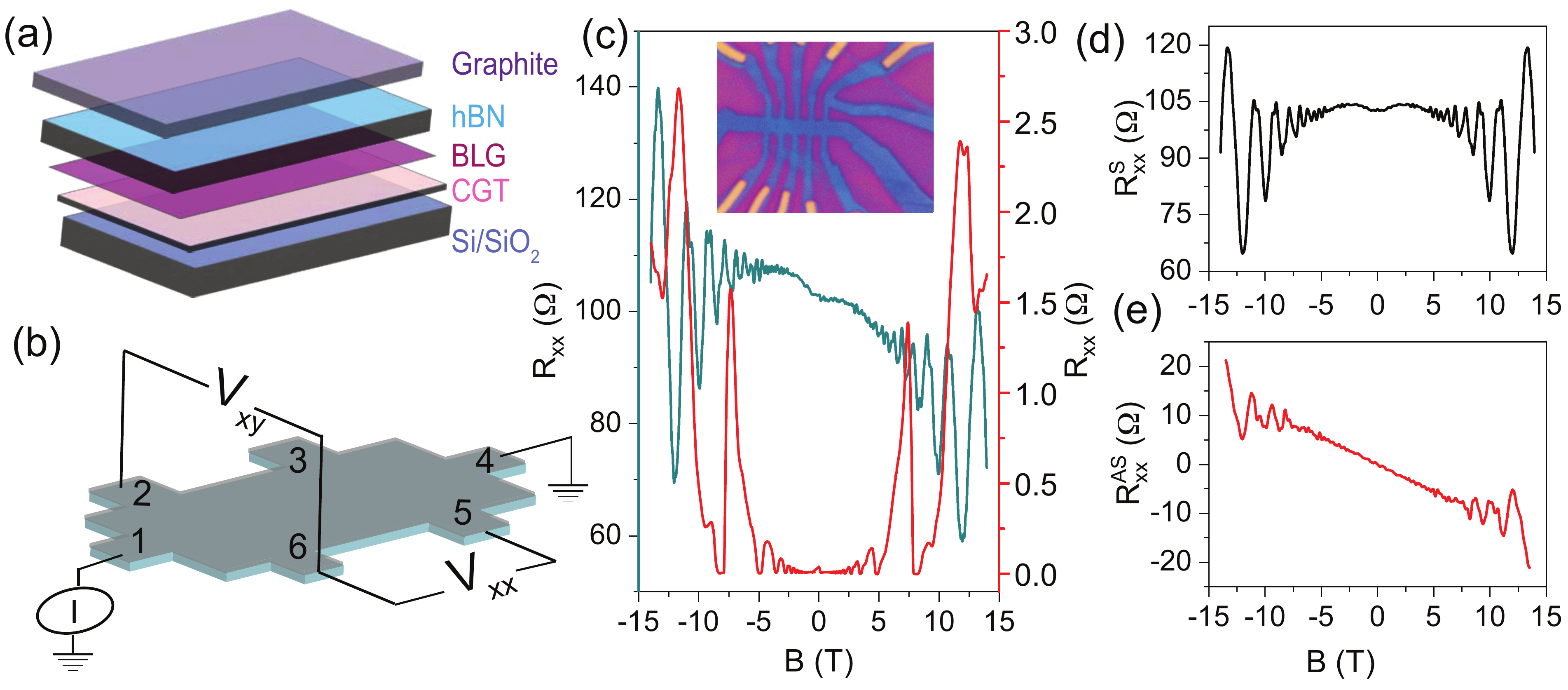}
			\small{\caption{\textbf{Device schematic and antisymmetric magnetoresistance:} (a) Schematic of the layers used to create the heterostructure. (b) Sketch of the device structure, including the contact numbers and the measurement configurations. (c) Comparison of the longitudinal magnetoresistance  ${R_{14,65}} = (V_{65}/I_{14})$ for a hBN/BLG/hBN heterostructure (red line; right-axis) and a hBN/BLG/CGT heterostructure (green line, left-axis). The data were measured at $2 ~{K}$. Inset: Optical image of the \textbf{D1}. (d) Plot of the symmetric component of the magnetoresistance  ${R^{\rm S}_{14,65} = [R_{14,65}(B)+R_{14,65}(-B)]/2}$ versus $B$ for the hBN/BLG/CGT device. (e) Plot of anti-symmetric component of the magnetoresistance  ${R^{\rm AS}_{14,65} = [R_{14,65}(B) - R_{14,65}(-B)]/2}$ versus ${ B}$ for the hBN/BLG/CGT device.}
		\label{fig:fig1}}
		\end{center}
 \end{figure*}

 \begin{figure*}[t!]
		\begin{center}
	\includegraphics[width=\columnwidth]{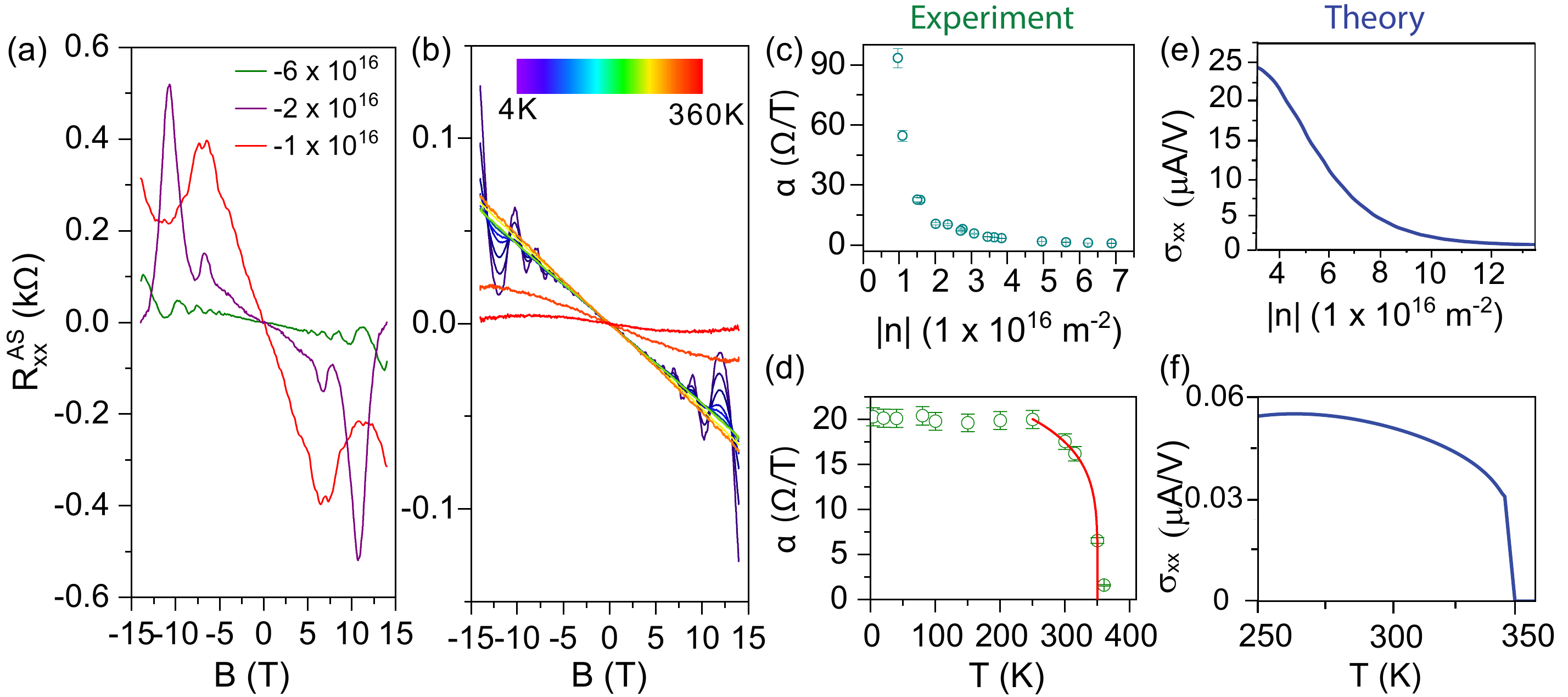}
	\small{\caption{\textbf {Number density and temperature dependence of anti-symmetric magnetoresistance:} (a) Plots of the ${R^{\rm AS}_{14,65}}$ versus $B$ at three representative number densities;  ${n=-6\times 10^{16}~\mathrm{m^{-2}}}$ (green curve),   $n=-2 \times 10^{16}~\mathrm{m^{-2}}$ (purple curve), and $n=-1 \times 10^{16}~\mathrm{m^{-2}}$ (red curve). The measurements were done at $T=2$~K. (b) Plots of  ${R^{\rm AS}_{14,65}}$ versus $B$ at a few representative temperatures between the $T = 2$~K and $T = 360$~K. The data were taken at  ${n=-2 \times 10^{16}~\mathrm{m^{-2}}}$. (c) Plot of $\alpha=(dR_{xx}/dB)$ versus $n$; the measurement was performed at $B=5$~T and $T=2$~K. (d) Plot of $\alpha$ versus $T$ measured at $B=5$~T and ${n=-2 \times 10^{16}~\mathrm{m^{-2}}}$ (green open symbols). The red line shows the fit to $\alpha=\alpha_0 (1-T_c/T)^\beta$. (e) Calculated antisymmetric component of the magnetoconductivity versus charge carrier density at $B=1$ T. (f) Calculated antisymmetric component of the magnetoconductivity versus $T$ at $B=1$ T. }
\label{fig:fig2}}
\end{center}
\end{figure*}

 \begin{figure*}[t!]
	\begin{center}
	\includegraphics[width=\columnwidth]{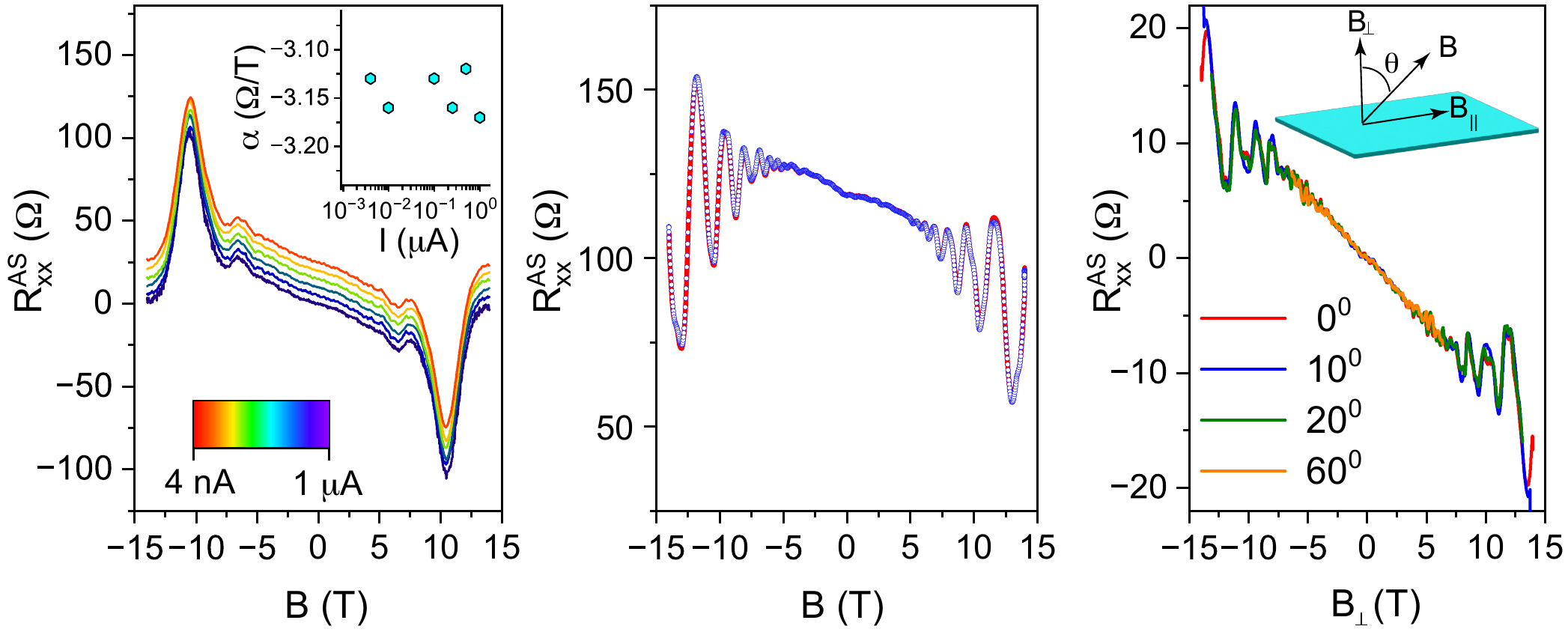}
 \small{\caption{\textbf{Current and parallel magnetic field dependence of antisymmetric magnetoresistance:} (a) Plot of ${R^{\rm AS}_{14,65}}$ for different bias currents ranging from $\mathrm{4n A-1\mu A}$. The data have been vertically offset for clarity. Inset: Plot of $\alpha$ versus $I$ showing the independence of the OMR on the magnitude of $I$. (b) Plots of ${R^{\rm AS}_{14,65}}$ (solid red line) and ${R^{\rm AS}_{41,56}}$ (open blue symbols). (c) The plot of ${R^{\rm AS}_{12,34}}$ versus $B_{\perp}$ for different values of the angle $\theta$ between the magnetic field and the perpendicular to the plane of the device (see inset of the figure). All the measurements were done at 2~K.  \label{fig:fig3}} }

    \end{center}
\end{figure*}

  \clearpage

 \begin{figure*}[t!]
		\begin{center}
			\includegraphics[width=\columnwidth]{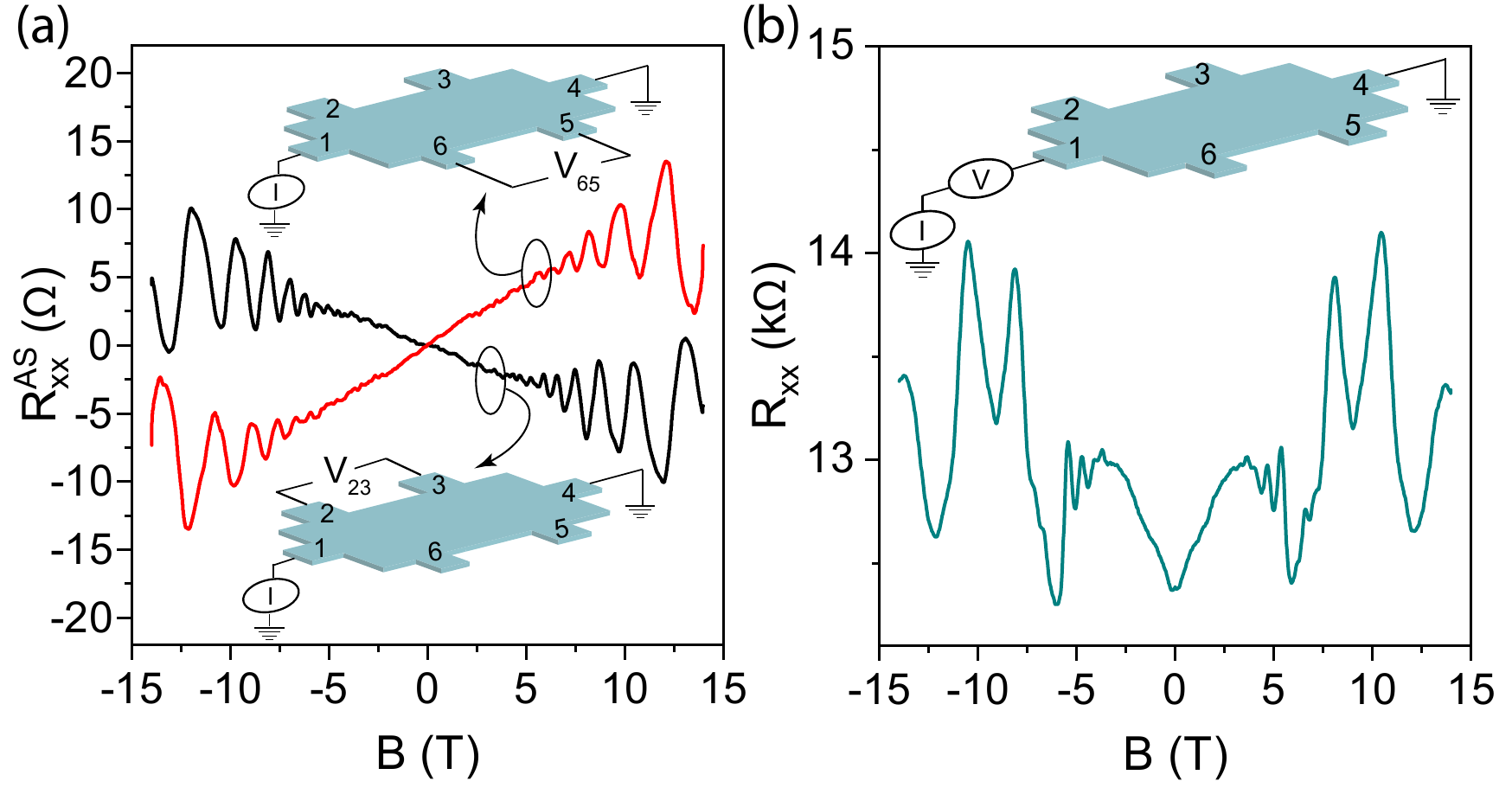}
   \small{\caption{\textbf {Edge property of antisymmetric magnetoresistance: }(a) OMR measured at the two opposite edges of the channel; ${R^{\rm AS}_{14,23}}$ (black curve) and ${R^{\rm AS}_{14, 65}}$ (red curve). (b) Plot of the 2-probe longitudinal magnetoresistance $R_{14,14}$ versus $B$. The 2-probe resistance does not have any antisymmetric component. The measurements were done at ${T=2}$~K, ${\theta = 0^\circ}$, and $n=-1 \times 10^{16} ~\mathrm{m^{-2}}$.
			\label{fig:fig4}	}}
		\end{center}
	\end{figure*}
 \clearpage

\section{Supplementary Information}

\renewcommand{\theequation}{S\arabic{equation}}
\renewcommand{\thesection}{S\arabic{section}}
\renewcommand{\thefigure}{S\arabic{figure}}
\renewcommand{\thetable}{S\arabic{table}}
\setcounter{table}{0}
\setcounter{figure}{0}
\setcounter{equation}{0}
\setcounter{section}{0}

\section{ Device fabrication}
Thin flakes of graphite, hBN, bilayer graphene (BLG), and \ch{Cr_2Ge_2Te_6} were mechanically exfoliated onto \ch{Si/SiO_2} substrates. To avoid degradation,  \ch{Cr_2Ge_2Te_6} was exfoliated inside an Argon-filled glove box. The thickness of the graphene flakes was initially estimated based on color contrast under an optical microscope and later confirmed using Raman spectroscopy. For \ch{Cr_2Ge_2Te_6} flakes, only optical contrast was used for thickness estimation; AFM measurement was avoided as few-layer \ch{Cr_2Ge_2Te_6} flakes are highly air-sensitive. A sequential pickup process was performed for the graphite, hBN, and BLG flakes using a polycarbonate (PC) film at $\mathrm{90^{\circ}C}$. The assembled heterostructure was transferred onto a \ch{Cr_2Ge_2Te_6} flake inside the glove box. The heterostructure was cleaned in chloroform, acetone, and IPA to remove the PC residue. Electron beam lithography was employed to define the contact and top gate electrodes. Reactive ion etching (mixture of \ch{CHF3} and \ch{O2} gas) was used to etch top hBN to make one-dimensional edge contacts to graphene. Cr/Pd/Au (3~nm/13~nm/55~nm) was deposited for making the electrical contacts, followed by liftoff in hot acetone and cleaning in IPA. The stack was finally dry-etched into a Hall bar. The top graphite flake is the top gate electrode, while the \ch{Si/SiO_2} substrate is the back gate.

\section{Gate voltage dependence of longitudinal resistance}

\begin{figure}[h]
	\centering
	\includegraphics[width=\linewidth]{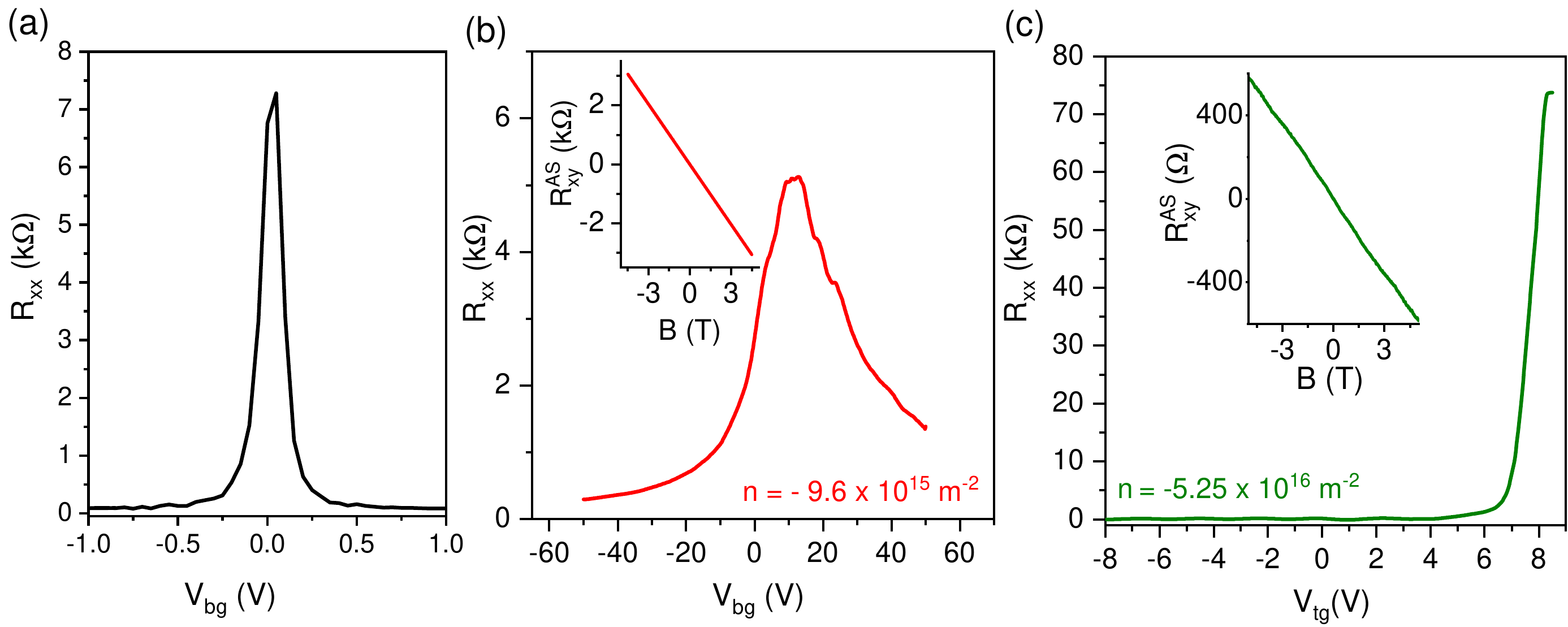}
	\caption{(a)  Longitudinal resistance versus gate voltage of  \ch{hBN/BLG/hBN} device. The charge neutrality point is at   $ n = 1.53 \times 10^{14}$~$\mathrm{m^{-2}}$ (b) Longitudinal resistance versus gate voltage of device \textbf{D1 }. The measurements are done at $ B=0$~T and $ T=2$~K. The inset shows the Hall resistance plotted against $ B$ measured data $ V_g = 0$~V, $ T=2$~K. The charge neutrality point is at $ n = 9.6 \times 10^{15}$~$\mathrm{m^{-2}}$. (c) Longitudinal resistance versus gate voltage of a few-layer \ch{Cr_2Ge_2Te_6}/BLG device. The measurements are done at $ B=0$~T and $ T=2$~K. The inset shows the Hall resistance plotted against $ B$ measured data $ V_g = 0$~V, $ T=2$~K. The extracted carrier density is  $ n = 5.3 \times 10^{16}$~$\mathrm{m^{-2}}$.}
	\label{fig:S3}
\end{figure}

Fig.~\ref{fig:S3}(a) Plot  of longitudinal resistance for the  device \textbf{D2} (\ch{hBN/BLG/hBN} ) device.  From the resistance versus gate voltage plot, we calculate that the charge neutrality point is at $n = 1.53\times 10^{14}$~$\mathrm{m^{-2}}$.   Fig.~\ref{fig:S3}(b) plots the longitudinal resistance for the device \textbf{D1} versus the back-gate voltage. The shift in the Dirac point to almost $10$~V suggests a charge transfer at the interface. From Hall measurements, we estimate this charge transfer to be $n = - 9.6 \times 10^{15}$~$\mathrm{m^{-2}}$ (see inset of Fig.~\ref{fig:S3}(a)). Fig.~\ref{fig:S3}(c) is the corresponding data for a few-layer \ch{Cr_2Ge_2Te_6}/BLG  device. Compared to the ultra-thin CGT, we observe a charge transfer to graphene by the few-layer CGT. The zero gate voltage Hall resistance yields this doping to be $n_0 = - 5.3 \times 10^{16}$~$\mathrm{m^{-2}}$ (inset of Fig.~\ref{fig:S3}(b)). A similar observation has been reported in graphene with other ferromagnets due to interfacial charge transfer~\cite{tseng2022gate,yang2023unconventional}.

\section{Comparison of zero field and field cool magnetoresistance}

Fig.~\ref{Fig:S1} compares the longitudinal magnetoresistance data for field-cooled (in a 14~T field) with the zero-field cooled data. We observed that field-cooled and zero-field-cooled data overlap, indicating that the induced magnetization in the graphene due to the bi-/tri-layer CGT is minimal and not strong enough to be manifested as hysteresis in magnetoresistance.

\begin{figure}
	\centering
	\includegraphics[width=0.5\linewidth]{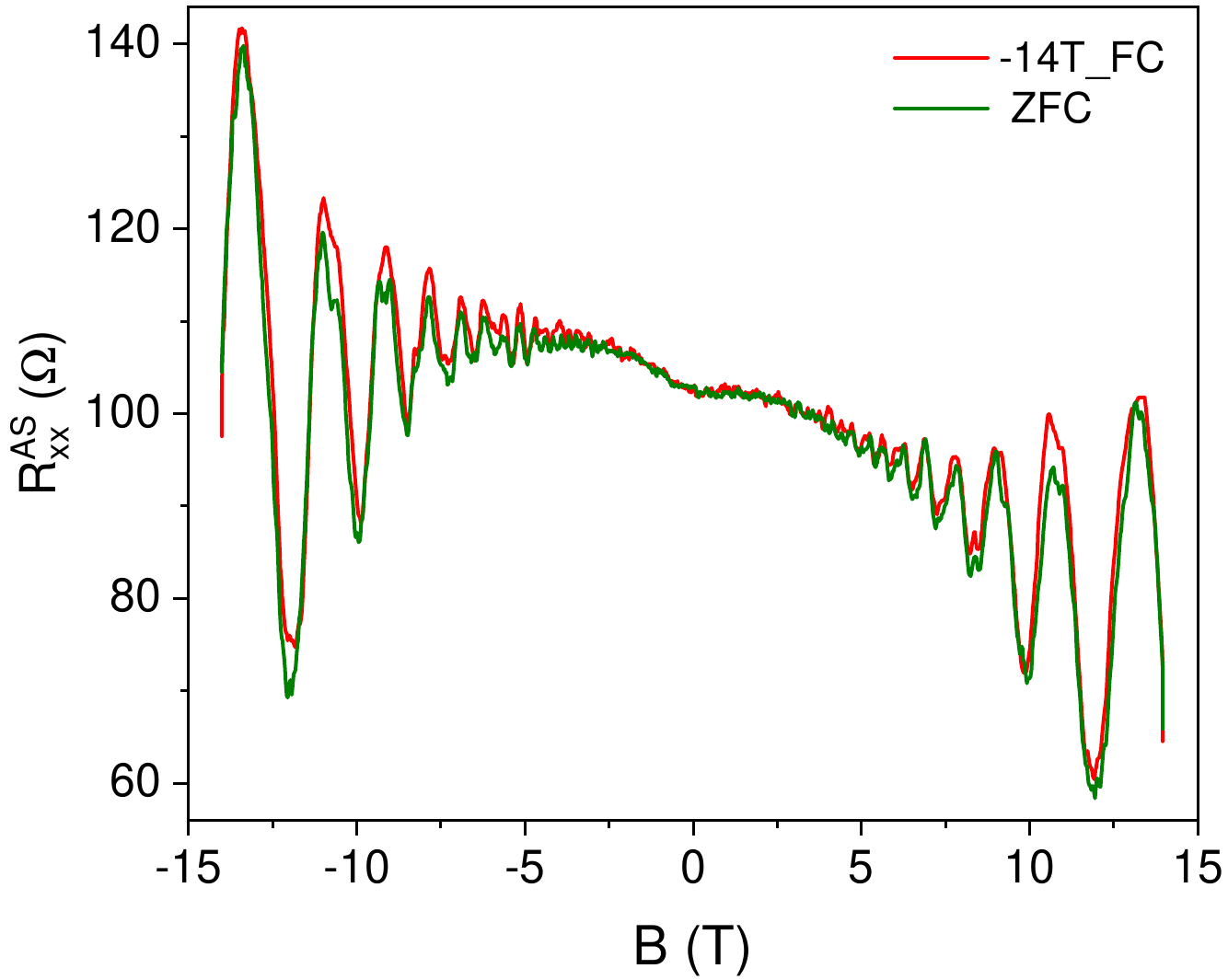}
	\caption{\textbf{Comparison of zero field cooled (green curve) and $-14$ T field cooled (red curve) OMR} . The data for field cool and zero field cool overlap, indicating a minimal magnetization in the heterostructure. The measurements were done  $ n = -6\times 10^{16}$~$\mathrm{m^{-2}}$ at $ T=2$~K.}
	\label{Fig:S1}
\end{figure}

\section{Large \% change in antisymmetric magnetoresistance}
We observe a remarkably large value of antisymmetric magnetoresistance reaching 40\% at ${n= -2.7 \times 10^{16}~ \mathrm{m^{-2}}}$ at $ T=100$~K. (Fig.~\ref{Fig:S2}). This value is much larger than the values of OMR reported in previous studies (Table~\ref{T1}).

\begin{table}[h!]
	\caption{Compilation of previous reports on odd parity magnetoresistance.}
	\centering
	\begin{tabular}{|c|l|c|c|} \hline
		Material&  Temperature & max. $ {R^{\rm AS}(\%) = 100\times\frac{[R(B) - R(-B)]}{2\times R(0)}} $ & Ref.\\ \hline
		Pyrochlore Iridate thin film$\mathrm{(Eu_2Ir_2O_7)}$&  2-105 K & 0.4\% (field-cooled) & \cite{fujita2015odd}\\ \hline
		InAs thin film & 2-300 K & 27\% & \cite{takiguchi2022giant}\\ \hline
		$\mathrm{SmCO_5}$& 300 K& 0.001\% & \cite{moubah2014antisymmetric}\\\hline
		$\mathrm{SmCO_5}$& 300K& 0.05\% & \cite{wang2020antisymmetric}\\\hline
		\textbf{BLG/CGT} & {\bf 300 K} & {\bf 40\%} & {\bf current study} \\\hline
	\end{tabular}

	\label{T1}
\end{table}
\begin{figure}
	\centering
	\includegraphics[width=0.5\linewidth]{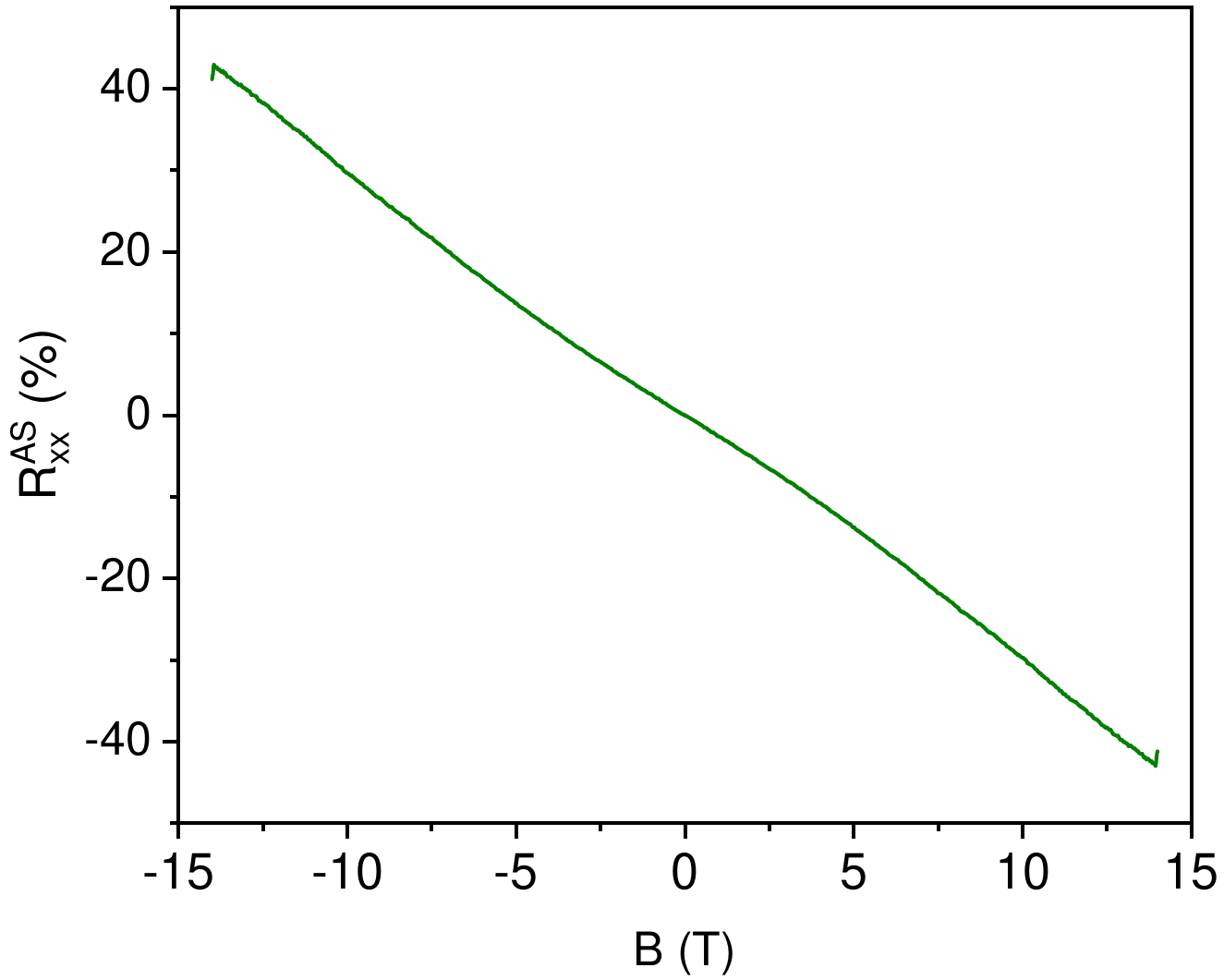}
	\caption{Plot of antisymmetric magnetoresistance  ${ R^{\rm AS}_{14,65}(\%) = 100\times[{ R}_{14,65}(B) - R_{14,65}(-B)]/2\times(R_{14,65}(0))}$ measured at $ n= -2.7 \times 10^{16}$~ $ \mathrm{m^{-2}}$ and $ T= 100$~K. }
	\label{Fig:S2}
\end{figure}

\section{Possible trivial origin of OMR: carrier density inhomogeneity picture}\label{Sec5}

Carrier density inhomogeneity in the transport channel can give rise to linear magnetoresistance. This effect has been studied extensively in high-mobility III-V semiconductors where, at high temperatures, carriers get excited to higher sub-bands, giving rise to multi-band transport~\cite{khouri2016linear}. This is not the case in bilayer graphene, where the higher energy bands cannot be accessed in our $T$ range of interest.

\begin{figure}
	\centering
	\includegraphics[width=0.5\linewidth]{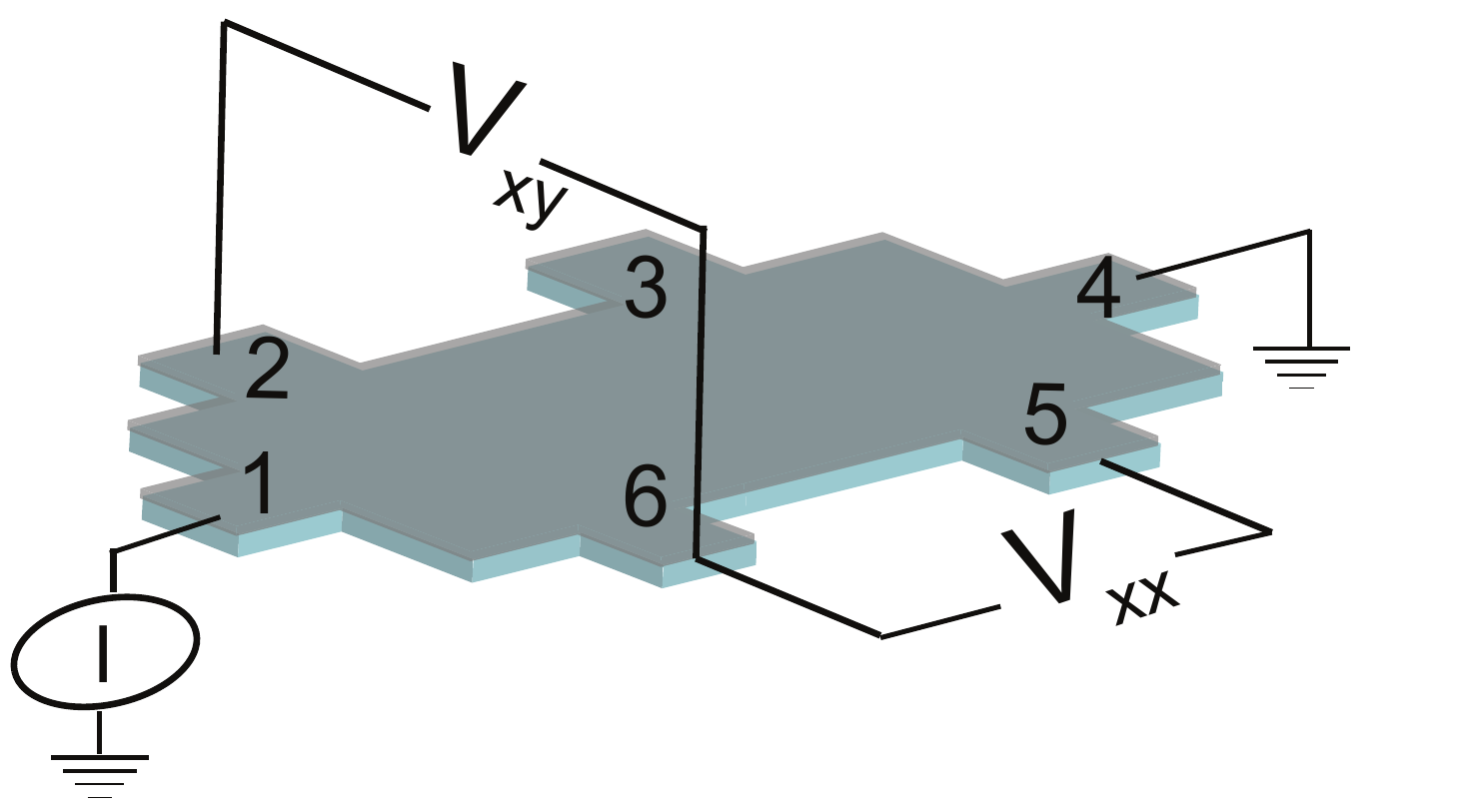}
	\caption{\textbf{Device configuration:} Sketch of the device  including the  contact probes and measurement configurations for longitudinal resistance $ (R_{14,65})$ and transverse resistance $ (R_{14,26}) $.}
	\label{fig:S4}
\end{figure}

Another trivial origin of OMR is number density inhomogeneity in the sample~\cite{PhysRevResearch.5.L032046}. We rule out this process in our BLG/CGT devices below. From Fig.~\ref{fig:S4}, we conclude that
\begin{equation} \label{S1}
	V_{26}-V_{35}= (V_2-V_6)-(V_3-V_5) = (V_2-V_3)-(V_6-V_5) = V_{23}-V_{65};
\end{equation}
The above equation implies that
\begin{equation} \label{S3}
	R_{14,26}^{\rm AS}-R_{14,35}^{\rm AS}=  R_{14,23}^{\rm AS}-R_{14,65}^{\rm AS}.
\end{equation}
Now,
\begin{equation}\label{S6}
	\begin{split}
		R_{14,26}^{\rm AS}-R_{14,35}^{\rm AS} & = -\frac{B}{e} \left[\frac{1}{n_{26}}-\frac{1}{n_{35}} \right]\\
		& = -\frac{B}{e} {\frac{\delta n}{n^2}}, \\
	\end{split}
\end{equation}
where $ \mathrm{ \delta n} = n_{35}-n_{26}$, $n_{35} (n_{26})$ is the average charge carrier density between the probes 3 and 5 (2 and 6)  and  $ {n= \sqrt{n_{35} n_{26}}}$~.
The above relations imply that the difference between the OMR measured between the two edges is:
\begin{equation}\label{S7}
	R_{14,23}^{\rm AS}-R_{14,65}^{\rm AS} = -\frac{B}{e} {\frac{\delta n}{n^2}}  .
\end{equation}

Using, $ R_{14,23}^{\rm AS} = -R_{14,65}^{\rm AS}$ (Fig.~4(a) of main manuscript), we finally obtain
\begin{equation}
	R_{14,23}^{\rm AS}\propto -\frac{B}{e} {\frac{\delta n}{n^2}}\\
\end{equation}
or,
\begin{equation}
	\alpha = \frac{R_{14,23}^{\rm AS}}{B} \propto \frac{1}{e} {\frac{|\delta n|}{n^2}}.
	\label{eq:S7}
\end{equation}
We observe a sharp twenty-fold decrease in $\alpha$ beyond 300~K while $ n$ increases by only a factor of two over this temperature range. Eq.~\eqref{eq:S7} then implies a sudden \textit{decrease} in the number density inhomogeneity by a factor of at least five with increasing $ T$ beyond $300$~K. Such a rapid decrease in $|\delta n|$ with increasing $T$ is unphysical; all measurements on graphene devices with moderate mobility to date show an increase in $|\delta n|$ as the device is heated.

We thus rule out number density variation along the graphene channel as the primary source of the observed OMR.

\section{Ruling out Hall intermixing as the origin of OMR: analysis of Hall and odd parity magnetoresistance for holes and electrons} \label{sec6}

Intermixing of Hall resistance in longitudinal magnetoresistance can also cause odd parity magnetoresistance. We can rule out this possibility based on the following observations. Fig.~\ref{fig:fig:S6}(a) shows the Hall resistance in the electron-doped region (solid red line) and for the hole-doped region (solid green line). As expected, $ R_{xy}$ for electrons and holes have opposite slopes. Fig.~\ref{fig:fig:S6}(b) shows the antisymmetric magnetoresistance for electrons (solid red line) and holes (solid green line) measured at the same number densities as in Fig.~\ref{fig:fig:S6}(a). By contrast to the Hall data, the two OMR plots for electron and hole doping have the same sign of the $\alpha = \mathrm{d} R^{\rm AS}_{xx}/\mathrm{d}B$. The identical response of odd parity magnetoresistance for both electrons and holes rules out the possibility that the origin of OMR is due to the intermixing of Hall resistance.

\begin{figure}
	\centering
	\includegraphics[width=1\linewidth]{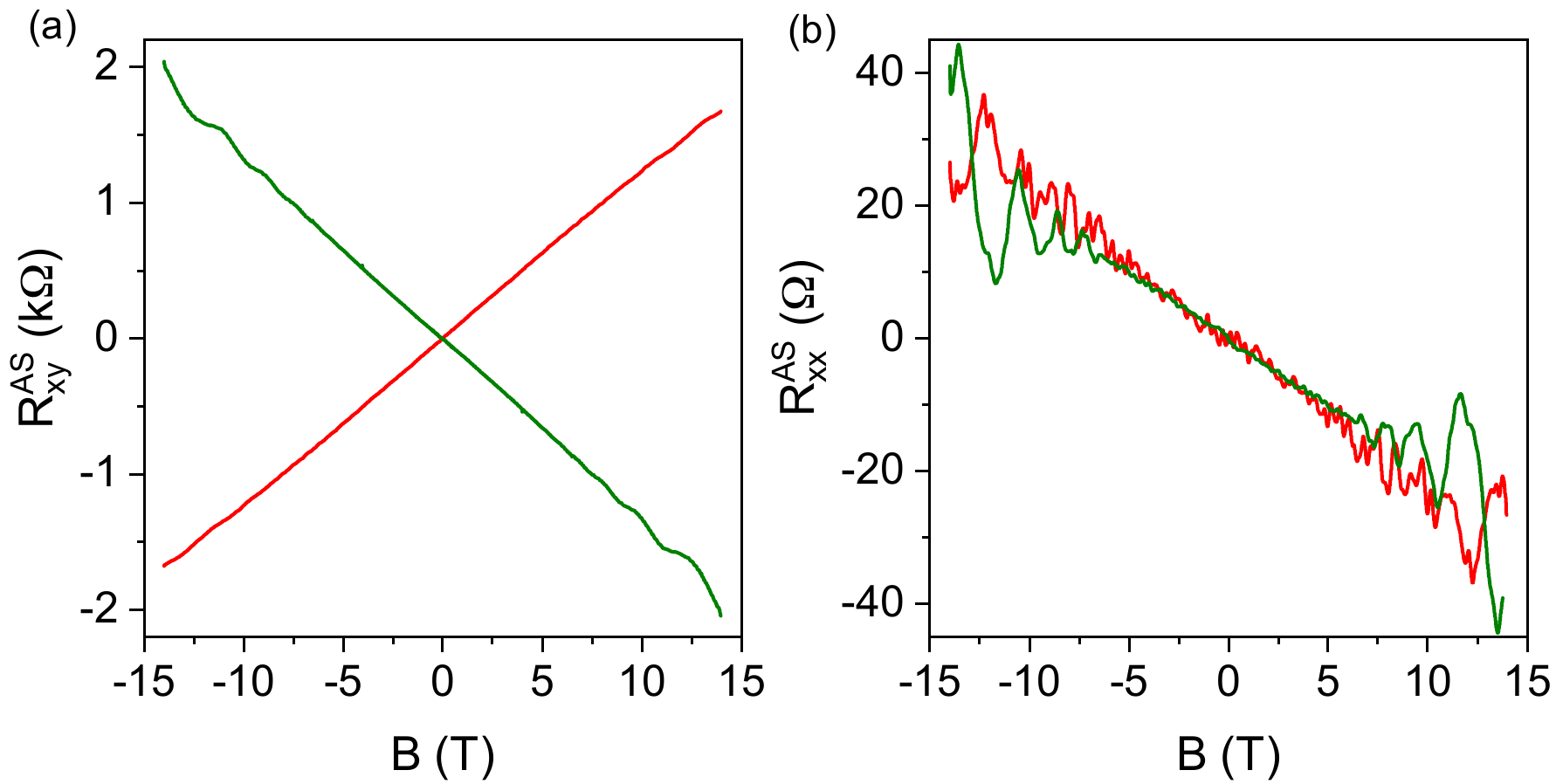}
	\caption{(a) Plot of Hall resistance for electrons (red) at carrier density of $ n = 4.97 \times 10^{16}$~$\mathrm{m^{-2}}$ and holes (green) at a carrier density  $ n = -4.84 \times 10^{16}$~$\mathrm{m^{-2}}$ at $ T=2$~K. (b) Antisymmetric magnetoresistance $ R_{xx}^{\rm AS}$ for electrons (red) and holes (green). While the hall resistance exhibits opposite slopes for electrons and holes, the antisymmetric magnetoresistance for both has the same slope. }
	\label{fig:fig:S6}
\end{figure}

\section{Band geometry induced antisymmetric magnetoresistance}
The measured antisymmetric magnetoresistance is linear in the magnetic field. Hence, to explain the antisymmetric magnetoresistance, we evaluate the expression of magneto-conductivity to order linear in the magnetic field. We use the semiclassical Boltzmann transport formalism to calculate the conductivity expressions. The equations of motions of electron's wave packets in the presence of an out-of-plane magnetic field are given by
\bea \label{eom_r}
\dot{\bm r}_p =   {\cal D} \left[\tilde{\bm v}_p +\frac{e}{\hbar}{\bm E}\times {\bm \Omega}\right],~~
\hbar\dot{\bm k }_p = {\cal D} \left[-e{\bm E} - e{ \tilde{\bm v}}_p \times {\bm B} \right]. \label{eom_k}
\eea
Here, `$-e$' is the electronic charge, $ \hbar \tilde{\bm v}_p={\bm \nabla}_{\bm k} \tilde{\varepsilon}_p = {\bm v}^0_p + {\bm v}^m_p$ is the band velocity for $p$-th band where $\tilde{\varepsilon} =\varepsilon_p-{\bm m}_p\cdot{\bm B}$ is the orbital magnetic moment modified energy. $\bm \Omega_p$ is the Berry curvature, which only has out-of-plane component for two-dimensional materials.
In Eq.~\eqref{eom_r}, ${\cal D} \equiv   1/(1 + \frac{e}{\hbar}{\bm \Omega}_p \cdot {\bm B})$ is the phase-space factor, which modifies the invariant phase-space volume according to $[d{\bm k}] \to [d{\bm k}] {\cal D}^{-1}$. The current density in the semiclassical formalism is given by $ {\bm j} = -e \sum_p \int \frac{d {\bm k}}{(2\pi)^2} {\cal D}^{-1} { \dot{\bm r}} g_p$, where $ g_p$ is the non-equilibrium distribution function.  The current density is related to the linear conductivity matrix via the relation $ j_a=\sigma_{ab} E_b$, with $a,b$ being the spatial indices $ (x,y)$. To obtain the $ g_p$ we solve the Boltzmann transport equation: $ \dfrac{\partial g_p}{\partial t} + \dot{\bm k } \cdot {\bm \nabla}_{\bm k}\,g_{p} = -\dfrac{g_p - f_p^{0}}{\tau} $. Here, $ f_p^0$ is the equilibrium Fermi-Dirac distribution function, and $\tau$ represents the scattering timescale. We obtain the non-equilibrium distribution function by solving the above equation in the steady state.
\be \label{distr_fn}
g_p =  f_{p}^0 +
\sum_{l=0}^{\infty} \left( \frac{e\tau}{\hbar} {\cal D}(\bm{\tilde v}_p \times \bm{B})\cdot\nabla_{\bm{k}} \right)^l \left[e \tau {\cal D} ( \bm{\tilde v}_p \cdot \bm{E} ) \frac{\partial f_p^0}{\partial \tilde \varepsilon_p} \right].
\ee
We mention that the $ l=1$ term gives rise to the conventional ordinary Hall effect due to the Lorentz force. We will set $ l=0$ since we are primarily interested in longitudinal magnetoconductivity, which is linear in the magnetic field. Using this, we find that the following expression gives the longitudinal magnetoconductivity to linear order in $ B$
\bea \label{sigma_B}
\sigma_{ab}(B) =&&  -2 e^2 \tau \int_{p,{\bm k}} {v}_{a}^m {v}_{b}^0 ~\partial_{\varepsilon} f^0 + \frac{e^3\tau}{\hbar} \int_{p,{\bm k}} ({\bm \Omega}\cdot {\bm B}) {v}_{a}^0 { v}_{b}^0 ~\partial_{\varepsilon} f^0 \nn \\
&&  + e^2 \tau \int_{p,{\bm k}} { v}_{a}^0 {v}_{b}^0  \left[ \partial_{\varepsilon}({\bm m}\cdot {\bm B}) ~\partial_{\varepsilon} f^0 + ({\bm m}\cdot {\bm B}) ~\partial^2_{\varepsilon} f^0 \right].
\eea
For brevity, we have denoted $ \int_{p,{\bm k}} \equiv \sum_p \int \frac{d{\bm k}}{(2\pi)^2} $, and we do not explicitly mention the band index in the quantities. We find that the $ \sigma_{ab}(B)$ is induced by the band geometric quantities, such as the out-of-plane components of the Berry curvature and the orbital magnetic moment. We emphasize that the above equation for the conductivity also gives rise to the transverse Hall component ($ a\neq b$), which is symmetric in spatial indices $ (a,b)$. This component is distinct from the Lorentz force-induced ordinary Hall and Berry curvature-induced anomalous Hall conductivity, which are antisymmetric in the spatial indices. Following Onsager's reciprocal relation, the linear in $ B$ conductivity of Eq.~\eqref{sigma_B} vanishes in systems with intrinsically time-reversal preserving systems.

The longitudinal and the transverse Hall resistivity is related to the conductivities via the relation; $ \rho_{xx} = \frac{\sigma_{xx}}{\sigma_{xx}^2 + \sigma_{xy}^2 }$, and $ \rho_{xy} = \frac{\sigma_{xy}}{\sigma_{xx}^2 + \sigma_{xy}^2 }$. Here, the $ \sigma_{xx}$ and $ \sigma_{xy}$ contain both the symmetric and antisymmetric in $ B$ components of the conductivity. Assuming the symmetric part of the $ \sigma_{xy}$ is much smaller than the anti-symmetric part {\it i.e.,} $ \sigma_{xy}^{\rm AS} \gg \sigma_{xy}^{\rm S}$ and the symmetric part of $ \sigma_{xx}$ is much greater than the antisymmetric part {\it i.e.,} $ \sigma_{xx}^{\rm S} \gg \sigma_{xx}^{\rm AS}$, we obtain $ \rho_{xx}^{\rm AS} \approx \frac{\sigma_{xx}^{\rm AS}}{\sigma_{xx}^{\rm S}} \rho_{xx}^{\rm S}$, with $ \rho_{xx}^{\rm S}\approx \frac{\sigma_{xx}^{\rm S}}{(\sigma_{xx}^{\rm S})^2 + (\sigma_{xx}^{\rm AS})^2}$. This shows that the antisymmetric longitudinal resistance is directly proportional to the antisymmetric part of the longitudinal conductivity.


To qualitatively explain our experimental observation of antisymmetric MR, we numerically evaluate the antisymmetric magnetoconductivity for the bilayer graphene model proximitized with a single layer of Cr$_2$Ge$_2$Te$_6$ (BLG/CGT). The Hamiltonian for BLG/CGT heterostructure is given by~\cite{Fabian_prb21}
\be \label{Ham_blg_cgt}
{\cal H}= {\cal H}_{ orb} + {\cal H}_{ SOC}  + {\cal H}_{ ex} + {\cal H}_{ R},
\ee
with
\be \label{H_orb}
{\cal H}_{ orb} = \begin{pmatrix}
	\Delta  & \gamma_0 f_k & \gamma_4 f^*_k & \gamma_1 \\
	\gamma_0 f^*_k & \Delta & \gamma_3 f_k & \gamma_4 f^*_k \\
	\gamma_4 f_k & \gamma_3 f^*_k & -\Delta & \gamma_0 f_k \\
	\gamma_1 & \gamma_4 f_k & \gamma_0 f^*_k & -\Delta
\end{pmatrix} \otimes s_0,
\ee
\be
{\cal H}_{ SOC}  + {\cal H}_{ ex} + {\cal H}_{ R} =  \begin{pmatrix}
	(\xi \lambda_I^{A1} - { M^{A1}})s_z & 2i \lambda_{R1} s^{\xi}_{-} &  0 & 0\\
	-2i \lambda_{ R1} s^{\xi}_{+} & (-\xi \lambda_I^{ B1} - { M^{B1}})s_z & 0 & 0\\
	0 & 0& (\xi \lambda_I^{ A2} - { M^{A2}})s_z & 2i \lambda_{ R2} s^{\xi}_{-} \\
	0 &0 & -2i \lambda_{ R2} s^{\xi}_{+} & (-\xi \lambda_I^{ B2} - { M^{B2}})s_z
\end{pmatrix}.
\ee
In Eq.~\eqref{H_orb}, $\gamma_j$ with $j=\{0,1,3,4\}$ describe the intra and interlayer hoppings in BLG. The $f_k= - \frac{\sqrt{3}a}{2}(\xi k_x - i k_y)$ around the Dirac points of valley $\xi$. The $s^\xi_\pm = \frac{1}{2} (s_x  \pm i \xi s_y)$, $s_0$ the $2\times2$ identity matrix and $s_{\{x,y,z\}}$ are the components of the Pauli matrix. The parameters $\lambda_I$ ($M$) describe the proximity-induced intrinsic SOC (exchange) of the corresponding layer and
sublattice atom ($C_{A1}, C_{B1}, C_{A2}, C_{B2}$). The
parameters $\lambda_{R1}$ and $\lambda_{R2}$ are the Rashba couplings of the two
individual graphene layers. Since our experimental setup consists of CGT on a single side (say, bottom) of BLG, we make the proximity-induced parameters for other (top) layers zero. We use the proximity parameters quoted in Ref.~\cite{Fabian_prb21} with $\Delta=5$ meV. The Berry curvature weighted dispersion relation for the Hamiltonian Eq.~\eqref{Ham_blg_cgt} is shown in Fig.~\ref{fig:fig7}. From Fig.~\ref{fig:fig7}, it is clear that as we move away from the Dirac point, the Berry curvature for the corresponding bands decreases rapidly. A similar feature can also be observed for the orbital magnetic moment. Since the OMR is induced by the Berry curvature and the OMM, which decreases with increasing energy vis-\'a-vis career density, the OMR strength decreases with the increasing number density.

\begin{figure}
	\centering
	\includegraphics[width=\linewidth]{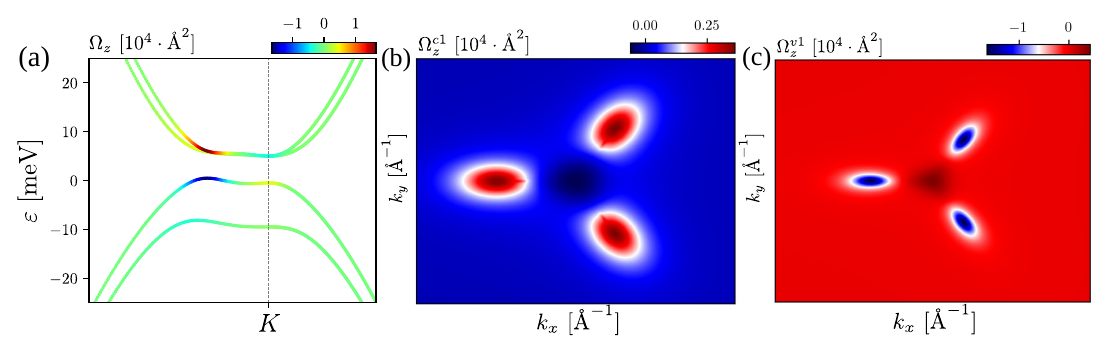}
	\caption{(a) The Berry curvature weighted band structure around one of the Dirac points ($K$) of bilayer graphene--single layer \ch{Cr2Ge2Te6} heterostructure. The Berry curvature $\Omega_z$ is only prominent in the narrow band gap region. The parameter used in this plot is taken from Ref.~\cite{Fabian_prb21}. The $\Omega_z$ distribution in $k_x-k_y$ space around $K$ point for the first conduction band (b) and valence band (c).}
	\label{fig:fig7}
\end{figure}

\section{A discussion on the Parish--Littlewood (PL) model, Abrikosov's quantum theory, and other theoretical models}

In this section, we briefly discuss some of the mechanisms of linear-$ B$-MR known in the literature and explain why these theories do not explain our results.

The PL model explains the OMR in a large inhomogeneous system~\cite{Parish2003}. In the presence of significant inhomogeneity in the system, the current flow gets distorted, and a part of the transverse Hall current appears in the longitudinal direction, which is linear in a magnetic field. This gives rise to the linear MR. However, our device quality is high, and our analysis shows no significant inhomogeneity, which can cause linear MR. The carrier density inhomogeneity can also cause the linear B MR, observed in semiconductors~\cite{khouri2016linear}. As discussed in Sections~\ref{Sec5} and \ref{sec6}, the inhomogeneity pictures can not explain OMR in our devices.

Abrikosov's quantum theory predicts linear-in-$ B$ MR for the systems in the extreme quantum limits~\cite{Abrikosov_prb98, Abrikosov2000}. Only the lowest Landau level should be filled in the extreme quantum limit, requiring very low carrier density and temperature. Our experimental observation of room temperature OMR down to a very low magnetic field regime rules out Abrikosov's quantum theory as a possible explanation for antisymmetric MR.

In Ref.~\cite{Song_prb15}, Song {\it et al.} proposed a semiclassical mechanism for linear-$B$-MR, where the guiding center motion dominates charge transport. Specifically, when the disorder potential varies on a scale that is large compared to the cyclotron radius, the Hall angle ($ \sigma_{xy}/\sigma_{xx}$) can become independent of the magnetic field. The longitudinal MR can be written as $ \rho_{xx}=\frac{\sigma_{xx}}{\sigma_{xx}^2 + \sigma_{xy}^2} = \frac{\cal G}{\sigma_{xy}}$ with ${\cal G}=\frac{\tan\theta_H}{1+\tan^2\theta_H}$, $ \tan\theta_H=\sigma_{xy}/\sigma_{xx}$. Consequently, when $ \tan\theta_H$ is independent of $B$, and using the relation $ \sigma_{xy} = ne /B$, we can see that the $ \rho_{xx}$ becomes linear in $ B$. As explained in Ref.~\cite{Song_prb15}, the theory only works for three-dimensional materials. Hence, this theory does not concern our experimental system and can not explain the observation of OMR.

Another theory for linear MR is the `hotspot' theory in the charge density wave phase of materials~\cite{Sinchenko_prb17, Wu_prb20, Feng_pnas19}. The electron scattering around the `hotspot' of the Fermi surface in the presence of CDW fluctuations leads to the linear field dependence of the scattering time. The field-dependent scattering finally gives rise to the linear-$B$-MR. However, no experimental evidence of a charge density wave phase in the graphene/CGT system has been found. Therefore, the `hotspot' theory of electron scattering can not explain our experimental results.

In Ref.~\cite{Wang_prb20anomalous}, a phenomenological model for linear MR was proposed based on linear band structure. It was argued that for linear bands, the change in spin-Zeeman energy modified density of states (at the Fermi level) between two spin channels with the magnetic field is linearly proportional to the applied magnetic field. This gives rise to the MR $\propto B$ in systems where electron-phonon scattering dominates (scattering time $\propto $ DOS$^{-1}$). In the bilayer graphene system, the low-energy bands are quadratic in the crystal momentum. Consequently, the DOS for the 2D system is independent of energy and the Zeeman magnetic field. Thus, Zeeman splitting modified linear band structure theory for linear-$ B$-MR fails to capture our experimental observations.

In Ref.~\cite{takiguchi2022giant}, the OMR in the 1D edge channel of InAs/(Ga, FeSb) heterostructure was explained to arise from the spin-Zeeman coupling. The magnetic field modifies the band energy via the spin-Zeeman coupling term. Asymmetric scattering between two helical spin-orbit coupled bands makes the Drude conductivity linearly magnetic field dependent. When the OMR is associated with the spin-Zeeman term, we expect it to be sensitive to the in-plane and out-of-plane magnetic field components. However, as mentioned earlier, the OMR in BLG/CGT is insensitive to the in-plane magnetic field component, which rules out the theory of OMR presented in Ref.~\cite{takiguchi2022giant} for our system.

We thus conclude that a topological band with a magnetically ordered state is the most plausible theory that qualitatively explains our result.

 \bibliography{arxiv}
\end{document}